\documentclass[pra,aps,nofootinbib,onecolumn,hyperref,showpacs,preprint]{revtex4}
\usepackage{graphics}
\usepackage{epsfig}
\usepackage{natbib}
\usepackage{amssymb}
\usepackage{bm,amsmath}

\begin{document}
\title{The dynamics of three-level $\Lambda$-type system driven by the trains of ultrashort laser pulses}

\author{Ekaterina Ilinova}
\affiliation{Department of Physics, University of Nevada, Reno,
Nevada 89557, USA}
\author{Andrei Derevianko}
\affiliation{Department of Physics, University of Nevada, Reno,
Nevada 89557, USA}

\begin{abstract}
We study the dynamics of a tree-level $\Lambda$-type atoms driven by a coherent train of short, non-overlapping laser pulses.
We derive analytical non-perturbative expressions for density matrix by approximating pulses by delta-function.
We demonstrate that depending on train parameters several scenarios of system dynamics are realized.
We show the possibility of driving Raman transitions between the two ground states of $\Lambda$-system avoiding populating excited state by using the pulses with
effective area equal to $2\pi$. The number of $2\pi$-pulses needed to transfer the entire population from one ground state to another depends on the ratio between the Rabi frequencies of two allowed transitions.
In the case of equal Rabi frequencies, the system can be transferred from one ground state to another with a single $2\pi$ pulse.
When the total pulse area differs from $2\pi$ and the two-photon resonance condition is fulfilled,
the system evolves into a ``dark'' state and becomes transparent to subsequent pulses.
The third possible scenario is the quasi-steady-state regime when neither the total single pulse area is equal to $2\pi$ nor the two-photon resonance condition is fullfilled. In this regime the  radiative-decay-induced drop in the population following a given pulse is fully restored by the subsequent pulse. We derive analytical expression for the density matrix in the quasi-steady-state regime.  We analyze the dependence of the post-pulse excited state population in the quasi-steady-state regime on the train parameters.  We find the optimal values for train parameters corresponding  to the maximimum of the excited state population. The maximum of the excited state population in the steady state regime is reached at the effective single pulse area equal to $\pi$ and is equal to $2/3$ in the limiting case when its radiative lifetime is much shorter then the pulse repetition period.
\end{abstract}


\maketitle
\section{Introduction}
\begin{figure}[h]
\begin{center}
\includegraphics*[width=5.2in]{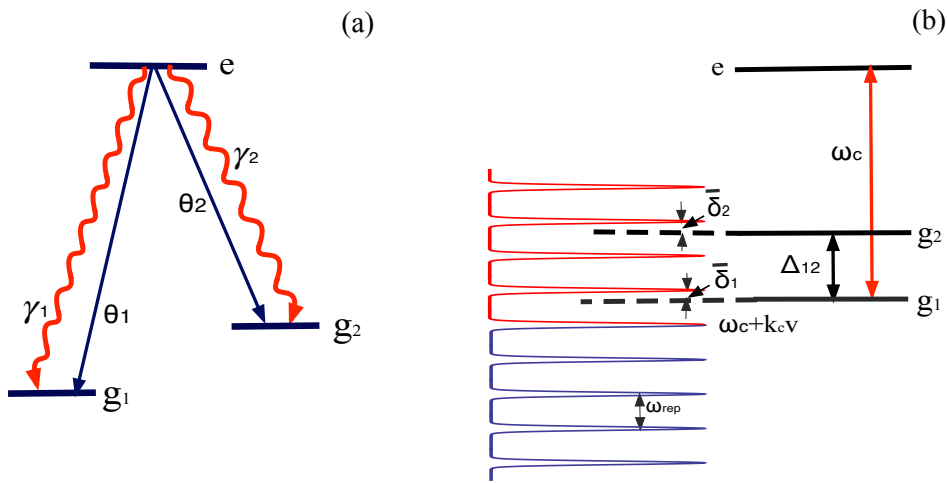}
\end{center}
\caption{ (color online ) Energy levels of $\Lambda$-system and positions of frequency comb teeth. The comb is Doppler shifted in the atomic frame moving with velocity $v$.  \label{Fig:lambda}}%
\label{Fig:Setup}%
\end{figure}

The frequency combs (FC) generated by the trains of ultrashort laser pulses \cite{UdeHolHan02} have been actively developed over the past 10 years. Recently a fiber-laser-based FC with 10 W average power was demonstrated \cite{SchHarYos08}  with the prospects of further scaling of technology up to 10 kW average power.  The spectral coverage of combs has been extended from optical to the ultra-violet and mid-IR spectral range \cite{AdlCosTho09,  LeiMarBye11,  VodSorSor11}.  High resolution quantum control via the combination of pulse shaping and frequency comb was shown  in \cite{Ye07,ZhiLeaWei07,StoCruFla06,StoMatAvi08}. The experiments on line-by-line addressing have been done \cite{ZhiLeaWei07}. These rapid technological developments enable novel applications in precise metrology \cite{UdeHolHan02,DerKat11},  atomic and molecular spectroscopy \cite{DiaHolMbe07},  quantum computing \cite{HayMatMau10,GarZolCir03}, manipulating  external and internal  degrees of freedom of atomic and molecular systems \cite{ShaPeeJun08,VitChoAll08,ShiMal10}.
In this paper we explore the dynamics of three-level $\Lambda$-type atoms (Fig.\,~\ref{Fig:lambda}(a)) interacting with coherent train of ultrashort laser pulses.
The dynamics of two-\cite{StrHorKer91,  AumBanSken05,  AllAri93,  IliAhmDer11},  three- \cite{FelAciVia04, ShiMal10,  FelBosAci03} and multi-level systems \cite{ShaPeeJun08} driven by such trains has  been actively investigated over the past decade.  In particular there were proposals for Doppler cooling of atoms based on two-photon transitions driven by ultrafast pulse trains\cite{Kie06},  optical pumping and vibrational cooling of molecules by femtosecond-shaped pulses \cite{VitChoAll08} and rotational cooling of molecules by chirped laser pulses \cite{ShiMal10}.

The analytical expression for the density matrix of a two-level system interacting with the pulse train was obtained in \cite{FelBosAci03,IliAhmDer11}
However, in many cases the atom can not be approximated as a two-level system because the excited state decay to intermediate sublevels.
As a practicle example of $\Lambda$ system, we consider the ground states of group III atoms. Their ground states are composed of two fine-structure sublevels $nP_{1/2}$ and $nP_{3/2}$ and the decay to the intermediate level can not be neglected \cite{PruAri03}.


Recently there was a series of works studying the dynamics of three-level atoms interacting with a train of ultrashort pulses\cite{FelAciVia04,SoaAra07,MorVia11}.

In particular, accumulative effects in the coherence of three-level atoms excited by femtosecond-laser frequency combs was studied in \cite{FelAciVia04}.
There authors obtained perturbative iterative solution for the density matrix of three-level system.
Coherent population trapping was studied in \cite{SoaAra07,MorVia11}. Perturbative analytical iterative solution for the density matrix in a weak field limit was presented.

Here we derive the analytical non-perturbative expressions for the density matrix of a $\Lambda$-system driven by coherent trains of ultrashort laser pulses.
Our work can be considered as extension of earlier works for a two-level system driven by the pulse train \cite{FelBosAci03,IliAhmDer11}.
As in our previous work  on two-level system \cite{IliAhmDer11}, we use the model of delta-function shaped pulses.  Using the derived equations we study dependence of system dynamics on the parameters of the pulse train.

For the pulse-train-driven $\Lambda$-system there are two major qualitative effects: ``memory'' and ``pathway-interference'' effects.  Both effects play important role in understanding of multilevel-system dynamics driven by the pulse train.

The system retains the memory of the preceding pulse as long as the population of the excited state does not decay between subsequent pulses. Then the quantum-mechanical amplitudes driven by successive pulses interfere and the spectral response of the system reflects the underlying frequency-comb structure of the pulse train. If we fix the atomic lifetime and increase the period between the pulses, the interference pattern is expected to ``wash out'', with a complete loss of memory in the limit  of large decay rates.  This memory effect is qualitatively identical to the case of the two-level system, explored in Ref.~\cite{IliAhmDer11}.

The ``pathway-interference'' effect is unique for  multilevel systems. The excited-state amplitude arises from simultaneous excitations of the two ground states. The two excitation pathways interfere. The ``pathway-interference'' effect is perhaps most dramatic  in the CPT regime \cite{Har97,MorVia11,SoaAra07,SoaAra10} where the ``dark'' superposition of the ground states conspires to interfere destructively, so that there is no population transfer to the excited state at all.

We show that in a particular case when the integer number of FC teeth fits into the energy gap between the two ground states and the total single pulse area differs from $2\pi$  the system evolves into a ``dark'' superposition of ground states and becomes transparent to the following pulses.  The ratio of the populations of two ground states in this regime is determined by the ratio of corresponding pulse areas. This effect is commonly referred to as a coherent population trapping (CPT) \cite{Har97,MorVia11,SoaAra07,SoaAra10}.

We also show that when the total single pulse area  (defined as the geometric sum of individual pulse areas $\theta_1$, $\theta_2$, corresponding to two different transitions, $\Theta=\sqrt{\theta_1^2+\theta_2^2}$ ) is equal to $2 \pi$  then Raman transitions can be driven between the two ground states avoiding the excited state. The number of pulses needed for complete population transfer from one ground state to another depends on the ratio between two pulse areas $\theta_1$, $\theta_2$. In a particular case of equal pulse areas the entire population can be transferred from one ground state to another by a single pulse.

Finally, we derive analytical expression for the density matrix of a system in a steady state regime realized for finite decay rate of the excited state and the total single pulse area $\Theta\neq 2\pi$. In this regime the  radiative-decay-induced drop in the population following a given pulse is fully restored by the subsequent pulse. We analyze the dependence of the  quasi-steady-state post-pulse excited state population on the FC  parameters.

This paper is organized as follows. In section I we derive general non-perturbatuve recurrent equation for
density matrix of $\Lambda$-system interacting with a coherent train of ultrashort laser pulses.  In section II we enumerate main parameters characterising interaction of $\Lambda$-system with a pulse train. In section III we study different scenarios of the system dynamics, each realized for certain combination of parameters. Finally, conclusions are drawn in Sec. \ref{Sec: Conclsn}.


\section{Analytical solution of the optical Bloch equations for a
delta-function pulse train}

\label{Sec:Propagators}
In a typical setup, a train of phase-coherent pulses is generated by
multiple reflections of a single pulse injected into an optical
cavity. A short pulse is outcoupled every roundtrip of the
wavepacket inside the cavity, determining a repetition time $T$
between subsequent pulses. At a fixed spatial coordinate, the
electric field of the train may be parameterized as

\begin{equation}
\mathbf{E}(t)=\hat{\varepsilon}\,E_{p}\,\sum\limits_{m}\cos(\omega_{c}%
t+\Phi_{m})\,g(t-mT)
\label{Eq:TrainField} \, ,
\end{equation}
where $\hat{\varepsilon}$ is the polarization vector, $E_{p}$ is
the field amplitude, and $\Phi_{m}$ is the phase shift. The
frequency $\omega_{c}$ is the carrier frequency of the laser field
and $g(t)$ is the shape of the pulses. We normalize $g(t)$ so that $\max |g( t)|
\equiv 1$, then $E_{p}$ has the meaning of the peak amplitude.
While typically pulses have identical shapes and
$\Phi_{m}=\Phi(mT)$, one may want to install an active optical
element at the output of the cavity that could vary the phase and
the shape of the pulses.

We are interested in a dynamics of three-level $\Lambda$-system,  interacting with the
train~(\ref{Eq:TrainField}), see Fig.\ref{Fig:lambda}. $\Lambda$-system is composed of the excited state
$|e\rangle$ and the ground states $|g_1\rangle$ $|g_2\rangle$ separated by $\Delta_{12}$; the transition frequencies between the excited and each of the ground states are $\omega_{eg_1}$, $\omega_{eg_2}$ correspondingly.
The optical Bloch equations (OBE) for the relevant density
matrix elements (populations $\rho_{ee}, \rho_{g_1g_1}, \rho_{g_2g_2}$ and coherences
$\rho_{eg_j}$ and $\rho_{g_je}$, $j=1,2$) read

\begin{eqnarray}\label{Eq:OBE1}
\dot{\rho}_{ee}&=&-\gamma\rho_{ee}-\sum\limits_{j=1}^2Im\left[\Omega_{eg_j} \rho_{eg_j}\right],\\
\dot{\rho}_{eg_j}&=&-\frac{\gamma}2\rho_{eg_j}+i \sum\limits_{j=1}^{2} \frac{\Omega^*_{eg_j}}2 (\rho_{ee}\delta_{jp}-\rho_{g_pg_j}), \label{Eq:OBE2}\\
\dot{\rho}_{g_jg_{j'}}&=&\delta_{jj'}\gamma_j \rho_{ee}+\frac{i}2(\Omega^*_{eg_{j'}}\rho_{g_je}-\Omega_{eg_j}\rho_{eg_{j'}}).\label{Eq:OBE3}
\end{eqnarray}

The time- and space-dependent Rabi frequency is
\begin{equation}\label{Eq:rabifreqsp}
\Omega_{eg_j}(z,t)=\Omega^{peak}_j \, \sum_{m=0}^{N-1} g(t+\frac{z}{c}-m T )e^{-i(k_cz(t)-\delta_jt-\Phi_m)}\, ,
\,
\end{equation}
where $\delta_j=\omega_{c}-\omega_{eg_j}$, $k_c=\omega_c/c$ and $z$ is the
atomic coordinate). The peak Rabi-frequency $\Omega^{peak}_j=\frac{E_{p}%
}{\hbar}\langle e|\mathbf{D}\cdot\hat{\varepsilon}|g\rangle$ is expressed in terms of the dipole matrix element. Eqs.~(\ref{Eq:OBE1}, \ref{Eq:OBE2}, \ref{Eq:OBE3}) were derived using the rotating wave approximation.
Notice that the energy gap between the two ground states can be expressed in terms of individual detunings: $\Delta_{12}=\delta_2-\delta_1$.

Notice that as long as the duration of the pulse is
much shorter than the excited state lifetime and the repetition time, the atomic system behaves
as if it were a subject to a perturbation by a series of
delta-function-like pulses. In this limit, the only relevant
parameter affecting the quantum-mechanical time evolution is the
effective area of the pulse
\begin{equation}
\theta_j=\Omega_{j}^{peak}\,\int\limits_{-\infty}^{\infty}g(t)dt \, ,
\end{equation}
and $\Omega^{peak}_jg(t) \rightarrow \theta_j \delta(t)$ in all the previous expressions.
As an illustration, we may consider a Gaussian-shaped pulse, $g(t)=e^{-t^{2}%
/2\tau_{p}^{2}}$. In the limit $\tau_{p}\ll T$, this pulse is
equivalent to a delta-function pulse
$\lim\limits_{\tau_{p}\rightarrow0}e^{-t^{2}/2\tau
_{p}^{2}}\rightarrow\sqrt{2\pi}\tau_{p} \, \delta(t)$, as both pulses
have the very same effective area $\theta_j$.

Now we turn to finding the solution of the
OBEs for a coherent train of delta-function pulses,
\begin{eqnarray}
\dot{\rho}_{ee}&=&-\gamma\rho_{ee}-\sum\limits_{m=0}^{N-1}\delta(t-mT)\sum\limits_{j=1}^2(\theta_{j} Im\left[e^{-i(k_cz(t)-\delta_jt-\Phi_m)}\rho_{eg_j}\right],\\
\dot{\rho}_{eg_j}&=&-\frac{\gamma}2\rho_{eg_j}+\frac{i}2 \sum\limits_{m=0}^{N-1}\delta(t-t_m)\sum\limits_{p=1}^{2} \theta_{p}e^{i(k_cz(t)-\delta_pt-\Phi_m)}(\rho_{ee}\delta_{jp}-\rho_{g_pg_j}), \\
\dot{\rho}_{g_jg_{j'}}&=&\delta_{jj'}\gamma_j \rho_{ee}+\frac{i}2\sum\limits_{m=0}^{N-1}\delta(t-mT)(\theta_{j'}e^{i(k_cz(t)-\delta_{j'}t-\Phi_m)}\rho_{g_je}-\theta_{j}e^{-i(k_cz(t)-\delta_jt-\Phi_m)}\rho_{eg_{j'}}).\label{Eq:DMdeltamodel}
\end{eqnarray}

We will distinguish between pre-pulse (left) and post-pulse
(right) elements of the density matrix, e.g., $\left(
\rho_{ee}^{m}\right)  _{l}$ and $\left(  \rho_{ee}^{m}\right)
_{r}$ are the values of the excited state population just before and just after the
$(m+1)^{\text{th}}$ pulse. Below we relate these values at each pulse
and between the pulses. Starting from given initial values of
$\rho$ and applying a recurrent procedure we may find $\rho$ at
later times.

Delta-function pulses cause abrupt changes in density matrix
elements at points $t_{m}=mT$. Between the pulses, however, the
dynamics is simple as it is determined by the spontaneous decay.

This leads to the following time evolution between the pulses
($mT<t<\left(m+1\right)T)$)

\begin{eqnarray}\label{decayeq}
\rho_{ee}(t) &=&\left(\rho_{ee}^m\right)_{r}e^{-\gamma t},\\
\rho_{eg_j}(t) &=&\left(\rho_{eg_j}^m\right)_{r}e^{-\frac{\gamma}2t},\\
\rho_{g_jg_{j'}}(t) &=&\left(\rho_{_jg_{j'}}^m \right)_r+\delta_{j,j'}\frac{\gamma_j}{\gamma} \left(\rho_{ee}^m\right)_r (1-e^{-\gamma t}).
\end{eqnarray}

Further, we may neglect the spontaneous decay \emph{during} the
pulse, since for a typical femtosecond pulse $\gamma\tau_{p}\ll1$.
Then the OBEs in time interval ($mT-\varepsilon<t<mT+\varepsilon$), $\varepsilon\rightarrow 0^+$,
may be recast in the form $\dot{\rho}=-i\delta(t-mT)\left[
\boldsymbol{a}_{m},\rho\right]  $, where $\left[
\boldsymbol{a}_{m},\rho\right]  $ is a commutator and the matrix
$\boldsymbol{a}_{m}$ reads:
\begin{equation}
\boldsymbol{a}_m=\frac12\left(\begin{matrix}0&\theta_{1}e^{i\eta_{1}(t)}&\theta_{2}e^{i\eta_{2}(t)}\\ \theta_{1}e^{-i\eta_{1}(t)}&0&0\\ \theta_{2}e^{-i\eta_{2}(t)}&0&0\end{matrix}\right).
\end{equation}
Here
\begin{equation}\eta_i(t)=k_cz-\delta_it-\Phi(t).\label{Eq:etafirst}\end{equation}

The matrix notation corresponds to the following enumeration scheme for matrix elements of $\rho$
\begin{equation}
\rho=\left(\begin{matrix}\rho_{ee}&\rho_{eg_1}&\rho_{eg_2}\\\rho_{g_1e}&\rho_{g_1g_1}&\rho_{g_1g_2}\\\rho_{g_2e}&\rho_{g_2g_1}&\rho_{g_2g_2}\end{matrix}\right).
\end{equation}
The exact analytical solution of this equation is
$\rho\left( t\right)  =U^{\dag}(t)\rho(mT-\varepsilon)U(t)\equiv
U^{\dag}(t)\left(
\rho^{m}\right)  _{l}U(t)$, \ where $U(t)=\hat{T} \exp\left[  i\boldsymbol{a}_{m}%
\int_{mT-\varepsilon}^{t}\delta(t^{\prime}-mT)dt^{\prime}\right]
$, with $\hat{T}$ being the time-ordering operator. Thus the
pre- and post-pulse elements of the density matrix are related by
\begin{equation}
\left(  \rho^{m}\right)
_{r}=\mathbf{A}_m~\left( \rho^{m}\right)
_{l} \mathbf{A}_m^\dag,
\label{Eq:Rhoaccross}%
\end{equation}
where
\begin{equation}\label{Eq:Am}
\mathbf{A_m}=\left(\begin{matrix}\cos\frac{\Theta}2&-i\sin\frac{\Theta}2\sin\chi e^{i\eta_1(t_m)}&-i\cos\chi\sin\frac{\Theta}2e^{i\eta_2(t_m)}\\-i\sin\frac{\Theta}2\sin\chi e^{-i\eta_1(t_m)}&\cos^2\chi+\cos\frac{\Theta}2\sin^2\chi&-\sin^2\frac{\Theta}4\sin(2\chi)e^{-i\Delta_{12}t_m}\\
-i\cos\chi\sin\frac{\Theta}2 e^{-i\eta_2(t_m)}&-\sin^2\frac{\Theta}4\sin(2\chi)e^{i\Delta_{12}t_m}&\sin^2\chi+\cos\frac{\Theta}2\cos^2\chi
\end{matrix}\right).
\end{equation}
Here $\Theta=\sqrt{\theta_1^2+\theta_2^2}$ is the total single pulse area defined as the geometric sum of individual single pulse areas of the two transitions and $\chi=\arctan{\left(\frac{\theta_1}{\theta_2}\right)}$  determines their ratio.
\begin{figure}[h]
\begin{center}
  \includegraphics*[width=2.2in]{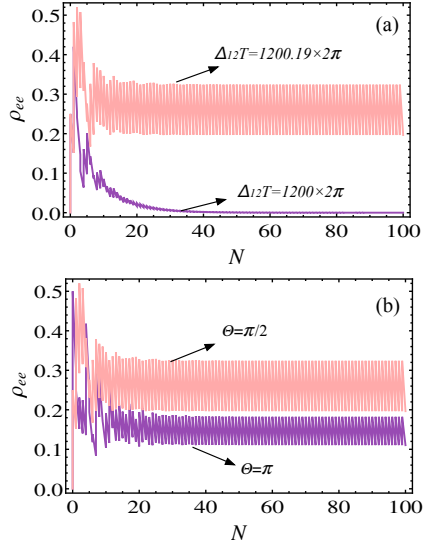}
\end{center}
\caption
{Time-evolution of the excited state population in
 $\Lambda$-system interacting with a coherent train of laser pulses.
The carrier frequency is resonant with the transition between the lowest ground and the excited state. The energy gap between the two ground states is $\Delta_{12}=2\pi \times 300 \, \mathrm{GHz}$ and
the decay rates are $\gamma_1=\gamma_2=2\pi\times 10 \mathrm{MHz}$ and $\theta_1=\theta_2$. (a) Comparison of two-photon-resonant (dark purple line - $\Delta_{12}T=1200\times2\pi$) and off-resonant (pink line - $\Delta_{12}T=1200.19\times2\pi$) regimes. The effective single pulse area is $\Theta=\frac{\pi}2$, $\theta_1=\theta_2$.  (b) The effect of the pulse area for a fixed value of $\Delta_{12}T=1200.19\times2\pi$: dark purple curve corresponds to $\Theta=\pi$ and the solution for $\Theta=\frac{\pi}2$ is shown in pink.
} \label{Fig:DMSS}\end{figure}
At this point, by combining Eq.~(\ref{decayeq}) and
Eq.~(\ref{Eq:Rhoaccross}) one may find time evolution of the
density matrix over a single repetition period; apparently, by
stacking these single-pulse and free-evolution propagators, one
may evolve a given initial $\rho$ over duration of the entire
train. In Fig.~\ref{Fig:DMSS} we show results of such calculation
for the excited state population of a heavy atom (atom remains at rest).

In Fig. \ref{Fig:DMSS} the atom is initially in the lowest ground state $|g_1\rangle$.
The lifetime of the excited atomic state is $15 \, \mathrm{ns}$, and the decay rates are equal: $\gamma_1=\gamma_2=\gamma/2$ .
If the frequency gap between the two ground states $\Delta_{12}$ is commensurate with the pulse repetition rate ($2\pi/T$) (see Fig.~\ref{Fig:DMSS} (a), dark purple line), then the system evolves into a ``dark'' superposition of ground states and becomes transparent to the pulses.
\section{Characteristic dimensionless parameters}

To streamline the  analysis we introduce dimensionless parameters, characterizing pulse-train cooling of the $\Lambda$-system.
\begin{enumerate}

\item[(i)]{ The ratio of the pulse repetition period and the lifetime of the excited state
\begin{equation}
 \mu=\gamma T \,.
 \end{equation}
This parameter will in particular characterize the spectral profile of the post-pulse excited state population.}
\item[(ii)]{ Single-pulse areas $\theta_j$ for the two transitions $|g_j\rangle\rightarrow|e\rangle$, $j=1,2$.
We will also employ two related auxiliary parameters:
the angle determining the ratio between the single pulse areas $\theta_1$, $\theta_2$
\begin{equation}\chi=\arctan{(\theta_1}/{\theta_2})\end{equation}

and the effective single-pulse area

\begin{equation}\Theta=\sqrt{\theta_1^2+\theta_2^2}.\end{equation}
}

\item[(iii)]{ Branching ratios, based on the decay rates of the excited state to the two ground states
\begin{equation} b_1=\gamma_1/\gamma, \quad b_2=\gamma_2/\gamma \, .
\end{equation}
Certainly $b_1+b_2=1$.
}
\item[(iv)]{ Number of teeth fitting in the energy gap $\hbar\Delta_{12}$ between the two ground states
\begin{equation} \kappa=\Delta_{12}/\omega_{rep} \, .\end{equation}
Notice that $\kappa$   generally is not an integer number. When it is integer, the two-photon resonance conditions are satisfied and the system evolves into the dark state.}

\item[(v)]{ Doppler shifted phase (\ref{Eq:etafirst}) offsets between subsequent pulses defined as
\begin{equation}
\overline{\eta}_1=\eta_1(t)-\eta_1(t+T)=(k_cv+\delta_1)T+\phi, \qquad \overline{\eta}_2= \overline{\eta}_1+2\pi\kappa \, .
\end{equation}
Here $v$ is the atomic velocity and $\phi$ is the carrier-envelope phase offset between subsequent pulses, i.e., $\phi=\Phi_{m+1} - \Phi_{m}$ in Eq.~(\ref{Eq:TrainField}). These phase parameters will be used to characterize the spectral profile of the excited state population. As shown below the density matrix of a system is a periodic function of $\overline{\eta}_1, \overline{\eta}_2$. The two phases are always related as
$$
 \overline{\eta}_2 - \overline{\eta}_1 = 2\pi\kappa\equiv2\pi\Delta_{12}/\omega_{rep} \,.
$$

}

\item[(vi)]{ Residual detunings  $\overline{\delta}_j$, $j=1,2$, between $|g_j\rangle$ levels and the nearest FC modes in the reference frame moving with the atom.
In general,
$\overline{\delta}_1=\overline{\eta}_1/T+2\pi n_1/T$ and  $\overline{\delta}_2=\overline{\eta}_2/T+2\pi n_2/T$, where integers $n_j=0,\pm 1..$ are chosen to renormalize the residual detunings to the interval
$-\omega_{rep}/2<\overline{\delta}_j<\omega_{rep}/2$. }

\end{enumerate} 
\section{System dynamics \label{QualitDescrip}}
Below we show that the system dynamics is mostly determined by four parameters $\kappa$, $\chi$, $\Theta$, $\gamma$. Depending on these parameters the following four scenarios may be realized. These different regimes are covered in individual subsections of this section.
\begin{enumerate}
\item[(a)]{ Dark state (CPT) regime is realized for
finite decay rate $\gamma$, when the integer number of FC teeth fits into the energy gap between the two ground states ($\kappa=0,1..$). Here the system evolves into a stationary superposition of two ground states (``dark'' state), which is transparent to the pulse train. }

\item[(b)]{ Stimulated Raman transitions between the two ground states (avoiding populating the excited state) are observed in the $\Lambda$-system when  the effective single pulse area is $\Theta=2\pi n$, $n=0,1..$ and the decay of the excited state within the pulse can be neglected ($\gamma \tau_p\ll 1$).  If initially  the system is in one of the ground states, then the excited state remains unpopulated after each new pulse and  the system evolves as a time-dependent superposition of two ground states $|g_j\rangle$.  Pulses lead to, discussed below,  abrupt change of coefficients in this superposition.  As shown below, at some special choice of $\chi$, the entire population can be transferred from one ground state to another by a single $\Theta=2\pi$ pulse.  The decay of the excited state can be neglcted for the number of pulses estimated as $N\approx 1/\gamma\tau_p$.}

 \item[(c)]{ If the lifetime of the excited state is much longer than the pulse repetition period $T$ then for a number of pulses, $N\ll1/(\gamma T)$, the dissipation can be neglected.  In this case, if the effective single pulse area is not a multiple of $2\pi$, ($\Theta\neq2\pi n$),  the population in  $\Lambda$-system oscillates between all three states. This is the transient regime preceding the quasi-steady-state regime. }

\item[(d)]{ The quasi-steady-state regime (QSS).  After $N\gg1/(\gamma \tau_p)\gg1/(\gamma T)$ pulses the system evolves into a saturated regime.
In this regime, the same fraction of population is driven to the excited state by each pulse, so the maximum value of $\left(\rho^s_{ee}\right)_s$ is reached at the moment of time just after each pulse. Between the pulses the excited state population exponentially decays to the ground states  and reaches its minium value just before the next pulse. These minimum and maximum values of the excited state population do not depend on the sequential number of the pulse. }
\end{enumerate}

\subsection{``Dark'' state (CPT)}
When the energy gap between the two ground states is commensurate with the distance between modes in a FC  ($\kappa=0,1,.$), the two-photon resonance condition is fulfilled \cite{Har97,MorVia11},
and (similar to the case of two CW sources) the Hamiltonian posesses stationary ``dark'' state. Here the atom is in a superposition of two ground states, described in the interaction picture by  the wave function
 \begin{equation}
 |\psi\rangle_{dark}=\cos(\chi)|g_1\rangle-\sin(\chi)|g_2\rangle.\label{Eq:dark state}
 \end{equation}
Once in the stationary state, the system dwells in it unless perturbed (e.g., pulse train parameters change). As a result, the system becomes transparent to the pulse train.
This can be also explained by the distructive interference between quantum probability amplitudes of the transitions $|g_j\rangle\leftrightarrows|e\rangle$ at  $\kappa=\Delta_{12}/\omega_{rep}$ for the system in a ``dark'' state.

The fact that the superposition (\ref{Eq:dark state}) is an eigenstate can be observed from the fact that the density matrix, corresponding to  (\ref{Eq:dark state}),
 \begin{equation}
 \rho^{dark}=\left(\begin{matrix}0&0&0\\0&\sin^2\chi&-
 \frac{\sin(2\chi)}2\\0&- \frac{\sin(2\chi)}2&\cos^2\chi\end{matrix}\right) \,.
 \end{equation}
commutes with the time-evolution operator $\mathbf{A_m}$ (\ref{Eq:Am}).
Notice that the dark state (\ref{Eq:dark state}) does not depend on the branching ratios $b_1, b_2$.
The two-photon resonance ($\kappa=0,1..$) is a prerequisite for the existence of a stationary state in $\Lambda$-system. Below we show that the ``dark'' state can be avoided for a large number of pulses ($N\sim1/(\gamma \tau_p)$) if the effective pulse area is $\Theta=2\pi n$, $n=1,2$.

\subsection{Stimulated Raman transitions between the two ground states}
When the effective single pulse area is $\Theta=2 \pi n$, $n=1..$, and the decay of the excited state during the pulse can be neglected ($\gamma \tau_p\ll1$), the system oscillates between the two ground states, avoiding populating the excited state altogether.  If the radiative decay during  each pulse can be neglected and the excited state population $\left(\rho_{ee}^s\right)_r$ is zero, then analytical expression for the  time-evolution operator after the $N$-th pulse can be obtained as the product $\mathbf{A}_N..\mathbf{A}_3\mathbf{A}_2\mathbf{A}_1$, where the operator $\mathbf{A}_m$ is defined by Eq. (\ref{Eq:Am}).  Knowing the time-evolution operator, one can express the wave function  (which initially was  in the lowest ground state)  after the $N$-th pulse as

\begin{equation}
\label{Eq:Psinondsp}
|\psi\rangle_{ndsp}^N=C_{g_1,N} |g_1\rangle +C_{g_2,N}|g_2\rangle,
 \end{equation}
where
\begin{eqnarray}
C_{g_1,N}&=&\frac{e^{iN\pi \kappa }}{\sin^2(2\chi)}\left( (-1)^N F_{aN}(\varphi)+F_{aN}(-\varphi)\right),\\
C_{g_2,N}&=&\frac{e^{-i(N-1)\pi\kappa} }{\sin(2\chi)}\left( (-1)^N F_{bN}(\varphi)-F_{bN}(-\varphi)\right),\\
F_{aN}(\varphi)&=&e^{iN\varphi}\frac{\left(\cos\varphi-\cot(\pi\kappa)\sin\varphi\right)^2}{1+\csc^2(2\chi)\left(\cos\varphi-\cot(\pi\kappa)\sin\varphi\right)^2},\\
F_{bN}(\varphi)&=&e^{iN\varphi}\frac{\left(\cos\varphi-\cot(\pi\kappa)\sin\varphi\right)}{1+\csc^2(2\chi)\left(\cos\varphi-\cot(\pi\kappa)\sin\varphi\right)^2},\\
\sin\varphi&=&\sin(\pi\kappa)\cos(2\chi)\qquad \cos\varphi>0.
\end{eqnarray}
In particular, when $\kappa=1/2+ 2n$ ($n=0,1..$), one has:
\begin{eqnarray}
C_{g_1,N}&=&e^{iN\pi \kappa }\left(\frac{1+(-1)^N}2\cos\left(N\left(\frac\pi2-2\chi\right)\right)-i\frac{1-(-1)^N}2\sin\left(N\left(\frac\pi2-2\chi\right)\right)\right)\nonumber,\\
C_{g_2,N}&=&e^{-i(N-1)\pi\kappa}  \left(i\frac{1+(-1)^N}2\sin\left(N\left(\frac\pi2-2\chi\right)\right)-\frac{1-(-1)^N}2\cos\left(N\left(\frac\pi2-2\chi\right)\right)\right)\nonumber.\\
\end{eqnarray}
The system which is initially in one ground state can be transfered to another ground state by $N=2k$ pulses when  $\chi=\frac\pi{8k}(2(k-l)-1)$ and by $N=2k-1$ pulses when $\chi=\frac\pi{4(2k-1)}(2(k-l)-1)$,  $k=1,2...$,  $l=0,..k-1$.  
In a special case of equal pulse areas  (for example, $\theta_1=\theta_2=\sqrt{2}\pi$, ($\chi=\pi/4$)) the entire population can be transferred from one ground state to another by a single $\Theta=2\pi$ pulse.
If initially the excited state was populated, then $\rho_{ee}$ either remains constant if there is no decay to the lower states or becomes distributed between the oscillating populations of the two ground states if there is a decay of excited state to any of the ground states.

It is worth highlighting the difference in meaning of the $2\pi$-pulse in two- and three-level systems.
In a two-level system the $2\pi$ pulse would drive the entire population to the excited state and then return to the ground state by the same pulse simultaneously. In the case of three-level system one could explain vanishing excited state population at the end of the $2\pi$ pulse (if it was zero before the pulse) in a similar fashion the same pulse drives the population to the upper state and then back to the superposition of the two ground states. The nature of this process is different from the well-known Stimulated Raman Adiabatic Passage (STIRAP) \cite{ShaPeeJun08}, involving two CW sources with slow-varying amplitudes and equal detunings between carrier frequencies and transition frequencies.
In our pulsed laser case driving the population between the two ground states avoiding the excited state is not affected by the difference in detunings $\overline{\delta}_1$, $\overline{\delta}_2$.
In the limiting case when the excited state is metastable $\gamma\rightarrow \infty$, the conclusions made here can be generalized for slow varying-envelope pulses as long as the conditions for $\Theta$ and $\chi$ remain fulfilled.

\subsection{Transient regime}
During initial sequence of  $N\ll1/(\gamma T)$ pulses the decay of the excited state can be neglected and the density matrix evolves as
\begin{equation}\label{Eq:DMNondissipative}
\rho_{transient}^N=\mathbf{A}_{N-1}..\mathbf{A}_1\mathbf{A}_0\rho_0 \mathbf{A}_0^\dag\mathbf{A}_1^\dag..\mathbf{A}_N^\dag, \end{equation}
where $\rho_0$ is the initial density matrix.
In this regime,  the $\Lambda$-system oscillates between all  three states.
At $\kappa=0,1..$ the  wave function describing the system after the $N$-th pulse (if initially all the population is in the ground state $|g_1\rangle$) can be expressed  as
\begin{eqnarray}\label{Eq:Psi3ndsp}
 |\psi\rangle_{transient}^N&=&C_{g_1}|g_1\rangle+C_{g_2}|g_2\rangle+C_e|e\rangle,\\
C_{e}&=&-i\sin\chi\sin\frac{N\Theta }{2},\\
C_{g_1}&=&\cos^2(\chi)+\sin^2(\chi)\cos\frac{N\Theta}{2},\\
C_{g_2}&=&-\sin(2\chi)\sin^2\frac{N\Theta}{4}.\\
\end{eqnarray}
During the transient regime the ``dark'' state is not reached yet even if the two-photon resonance condition is fulfilled \cite{SoaAra07}.

\subsection{Quasi-steady-state regime \label{SubSec:QSS}}

Similar to the case of two kicked coupled damped pendula \cite{HemPre88},  the system eventually reaches saturated regime, where the  radiative-decay-induced drop in the population following a given pulse is fully restored by the subsequent pulse.
We will refer to this behavior as the quasi-steady-state (QSS) regime.  As shown in the Appendix, the QSS regime allows for a fully analytical solution.
Since $\rho_{ee}\left(  t\right) =\rho_{ee}\left(
t+nT\right)$, in the QSS regime,  pre- and post-pulse values
$\left(  \rho_{ee}^{m}\right)_{l,r}$ do not depend on the pulse
number $m$ and we denote these values as $\left(
\rho_{ee}^{s}\right)
_{l,r}$. Furthermore, because of the radiative decay, $\left(  \rho_{ee}%
^{s}\right)  _{l}=e^{-\gamma T}\left(  \rho_{ee}^{s}\right)  _{r}$.
The general solution for the density matrix in the QSS regime can be obtained from a system of linear algebraic equations derived from condition: $\left(\check{\rho}^N\right)_r=\left(\check{\rho}^{N+1}\right)_r$,  where $(\check{\rho}^N)$ is the re-normalized density matrix (see the Appendix for details).
The solution is fully analytical, however it is unwieldy and here we present its simplified form obtained for equal pulse areas $\theta_{1}=\theta_{2}$ $(\chi=\pi/4, \Theta=\sqrt{2}\theta_1)$.

Solution for the arbitrary pulse area is given in Appendix.

The post-pulse value is
\begin{equation}
\left(\rho_{ee}^s\right)_r=\frac{2 e^{\frac{\gamma T}{2}}}D\sin ^2\left(\pi\kappa\right) \sin^2\frac{\Theta}2,\label{Eq:DMQSSbranch}
\end{equation}
where
\begin{eqnarray}\label{Eq:D}
D&=&\left(b_1 \cos\overline{\eta}_1+b_2 \cos\overline{\eta}_2\right) \left(4 \cos
\frac{\Theta}2-\sin^2\frac{\Theta}2-2\cos(2\pi\kappa)\cos^4\frac{\Theta
}{4}\right)\nonumber\\ & & -2 \left(b_2 \cos\overline{\eta}_1+b_1\cos\overline{\eta}_2\right)
\left(\sin ^4\frac{\Theta }{4}+\cos
^2\frac{\Theta}2\right)\nonumber \\
&&+2\sin(2\pi\kappa)(b_2\sin\overline{\eta}_2+b_1 \sin\overline{\eta}_1)\cos^4\frac{\Theta
}{4}-\nonumber \\&&-2 \cosh\frac{\gamma T}{2} \left(\sin^4\frac{\Theta }{4}+\cos ^2\frac{\Theta }{4}
\sin ^2\left(\pi\kappa\right)\right).
\end{eqnarray}

The equation (\ref{Eq:D}) is symmetric with respect to the swap of the ground state labels.
The dependence on the phase offset $\overline{\eta}_1$ (after substituting $\overline{\eta}_2=\overline{\eta}_1+2\pi\kappa$) is the result of interference between the elementary responses of a system to subsequent pulses (the persistent ``memory'' of the system).
Particularly, when  $\gamma T\rightarrow\infty$, the excited state completely decays between the pulses and the interference factor vanishes (the ``memory'' is erased),
\begin{equation}
\left(\rho_{ee}^s\right)_r\rightarrow  \frac{ 4\sin^2\left(\pi\kappa\right)}{\tan^2\frac{\Theta }{4}+\frac{\sin^2\left(\pi\kappa\right)}{\sin^2\frac{\Theta }{4}}}.
\label{Eq:DMQSSbranchlargedecay}
\end{equation}

At equal branching ratios $b_1=b_2=1/2$ the equation (\ref{Eq:DMQSSbranch}) can be simplified further

\begin{eqnarray}
\left(\rho_{ee}^{s}\right)_r&=&\frac{e^{\frac{\gamma T}{2}} \sin ^2\left(\pi\kappa\right) \sin^2\frac{\Theta}2}{4D'}, \nonumber\\
D'&=& \left(\cos \left(\pi\kappa\right) \cos
   \left(\bar{\eta}_1+\pi\kappa\right) \left(\cos ^2\left(\pi\kappa\right) \cos^4\left(\frac{\Theta }{4}\right)-\cos\frac{\Theta}2
   \right)+\right.\nonumber\\
  && \left.\cosh \left(\frac{\gamma T }{2}\right) \left(\sin^4\left(\frac{\Theta }{4}\right)+\cos^2\left(\frac{\Theta }{4}\right) \sin ^2\left(\pi\kappa\right)\right)\right).   \label{Eq:DMQSSsimplea}
\end{eqnarray}

\begin{figure}[h]
\begin{center}
  \includegraphics*[width=2.5in]{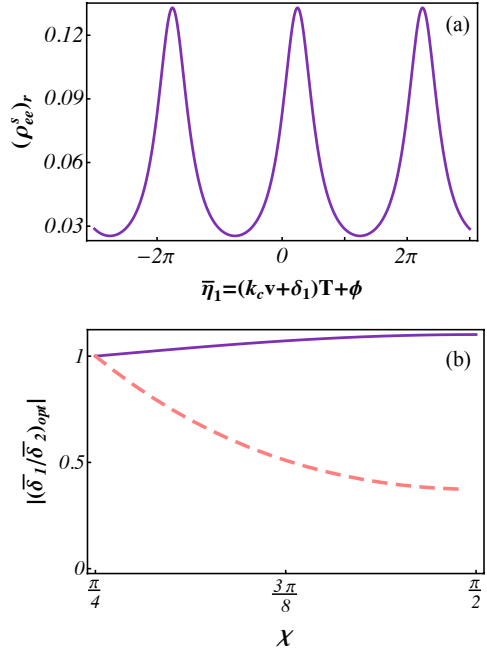}
\end{center}
\caption
{(a) The dependence of the QSS excited state population on the phase offset parameter $\overline{\eta}_1$ at fixed values of parameters $\gamma T=1/4$, $\Theta=\pi/3$, $b_1=1/2$, $b_2=1/2$ and $\kappa=0.12$.
(b) The dependence of the optimal ratio of the residual detunings $|(\bar{\delta}_1/\bar{\delta}_2)_{opt}|$ on the parameter $\chi$, corresponding to the maximum value of the post-pulse excited state population. The solid purple curve is obtained for equal branching ratios ($b_1=b_2$). The optimal ratio $|(\bar{\delta}_1/\bar{\delta}_2)_{opt}|$ for the case when the branching ratios are varied with the parameter $\chi$ as $b_1=\sin^2\chi=\theta_1^2/\Theta^2$, $b_2=\cos^2\chi=\theta_2^2/\Theta^2$ is shown with the dashed pink curve.} \label{Fig:QSSDMA}\end{figure}

We start analyzing the expression (\ref{Eq:DMQSSsimplea}) for the QSS value of the excited state population by studying its spectral profile. In Fig.~\ref{Fig:QSSDMA}(a) we plot the dependence of the excited state population $\left(\rho_{ee}^{s}\right)_r$ on the phase offset $\overline{\eta}_1$.  As an illustration we choose the following set of  parameters:  $\kappa=0.12+2n$ ($n=0,1..$), $b_1=b_2=1/2$, $\gamma T=1/4$, $\Theta=\pi/3$ (for example,  $\gamma=2\pi\times 10\, \mathrm{MHz}$, $T= 3.9804 \, \mathrm{ns}$, $\Delta_{12}=2\pi\times 300\, GHz $).

The periodic structure mimics the FC spectrum.
The maxima of $\left(\rho_{ee}^{s}\right)_r$  are reached at $\overline{\eta}_1=-mod(\pi\kappa,2\pi)+2\pi n, n=0,\pm1..$, that is at symmetrical values of detunings $\left(\overline{\delta}_1\right)_{opt}=-\left(\overline{\delta}_2\right)_{opt}=-mod(\pi\kappa,2\pi)/T$.

The optimal ratio of the residual detunings $\left(\bar{\delta}_1/\bar{\delta}_2\right)_{opt}$ can be defined as the value corresponding to the maximum post-pulse excited state population. At equal pulse areas $\theta_1=\theta_2$ and equal branching ratios $b_1=b_2=1/2$ this optimal value is equal to $-1$.
Generally, it depends on the ratio between branching ratios and the ratio between individual pulse areas $\theta_1/\theta_2$.
In case when one of the frequencies $\omega_{eg_j}$ is nearly resonant with the nearby FC tooth and another is not resonant with any of the FC modes, one would observe accumulation of the population in the state which is ``less coupled'' (not resonant).
This is the reason why at equal pulse areas ($\theta_1=\theta_2$) the accumulation of population in one of the ground states is reduced by setting the detunings $\overline{\delta}_1=-\overline{\delta}_2=-mod(\pi\kappa,2\pi)/T$. The detunings have to be of opposite signs to avoid the two-photon resonance which would drive the system into the ``dark'' state.

At different pulse areas and equal branching ratios $b_1=b_2=1/2$ the decay rates of the excited state to both ground states are equal but the  rate of the repumping of population from a certain ground state $|g_j\rangle$ to the excited state depends on the pulse area $\theta_j$  and the residual detuning $\bar{\delta}_j$.  When $\bar{\delta}_1=-\bar{\delta}_2$,  one would expect that the state corresponding to smaller pulse area $\theta_j$ accumulates the population and the excited-state population vanishes.  To mitigate this effect, the state of smaller effective pulse area $\theta_j$ has to be closer to the resonance with FC tooth (smaller detuning $\bar{\delta}_j$) than another one, corresponding to the larger effective pulse area.  Therefore,  one would expect that at different pulse areas and equal branching ratios, $b_1=b_2=1/2$, the optimal ratio between the detunings $\left(\overline{\delta}_1/\overline{\delta}_2\right)_{opt}$ at which the post-pulse excited state population has its maximum value, grows with increase of the parameter $\tan{\chi}=\theta_1/\theta_2$.
It is worth noticing  that at $\Delta_{12}\ll\omega_{eg_1}$ equal branching ratios mean equal dipole matrix elements, entering the definition of  Rabi frequency (\ref{Eq:rabifreqsp}). In order to obtain different pulse areas $\theta_{1,2}$ at equal branching ratios $b_1=b_2$, one would need additional  pulse shaping, to modulate intensities of  different FC teeth.

When the pulse areas scale proportionally to the square roots of branching ratios (all the teeth have the same intensity)  and $b_1\neq b_2$, the situation is different.
In this case, the ground state of  smaller pulse area is ``less coupled'' to the excited state, but the decay rate of the excited state to  this ground state is slower.  We found  that  in this case  the optimal ratio $\left(\bar{\delta}_1/\bar{\delta}_2\right)_{opt}$ decreases when increasing the ratio $\tan\chi=\frac{\theta_1}{\theta_2}$.
In Fig.~\ref{Fig:QSSDMA}(b) we plot the dependence of the ratio $\left(\overline{\delta}_1/\overline{\delta}_2\right)_{opt}$  as a function of the parameter $\chi$ at different values of branching ratios $b_1, b_2$.
The solid purple curve is obtained for equal branching ratios, $b_1=b_2=1/2$. The dashed pink curve was drawn assuming that the branching ratios vary with the parameter $\chi$ as $b_1=\sin^2\chi$, $b_2=\cos^2\chi$.

\begin{figure}[h]
\begin{center}
\includegraphics*[width=2.5in]{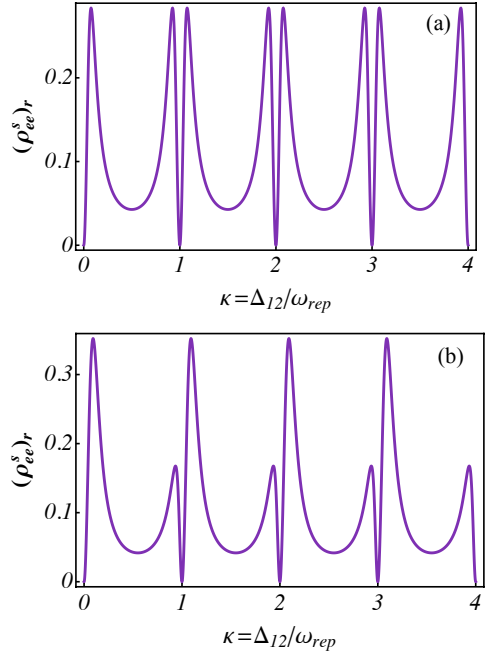}
\end{center}
\caption
{The dependence of the quasi-steady-state value of post-pulse excited state population on parameter $\kappa=\Delta_{12}/\omega_{rep}$ at $b_1=b_2$, $\gamma T=1/4$, $\Theta=\pi/3$. Panel (a) $\overline{\eta}_1=2\pi n$. Panel (b) $\overline{\eta}_1=-\pi/10\pm2\pi n, n=0,1..$.
} \label{Fig:QSSDM}\end{figure}

Next we study the dependence of the post-pulse excited state population value on the parameter $\kappa$, i.e., the ratio between the ground states energy gap and the pulse repetition frequency $\omega_{rep}$.
In Fig.~\ref{Fig:QSSDM} (a,b) we plot the dependence of the excited state population $\left(\rho_{ee}^{s}\right)_r$ (\ref{Eq:DMQSSsimplea}) on the parameter $\kappa=\Delta_{12}/\omega_{rep}$ at different values of $\overline{\eta}_1$.
Both curves exhibit periodic pattern which mimics the periodic spectrum of the pulse train.
The dips at integer values of $\kappa$ correspond to the CPT regime with zero excited state population.
Manipulating the pulse repetition rate $\omega_{rep}$ stretches the positions of FC modes in the frequency domain and consequently the residual detunings $\overline{\delta}_j$, between the frequencies $\omega_{eg_j}$ and nearest teeth.

The maxima of population in Fig. \ref{Fig:QSSDM} (a,b) remain the same when increasing the value of $\Delta_{12}/\omega_{rep}$ (that is increasing the value of $T$). This can be explained by the fact that at fixed value of the parameter $\gamma T$, the value of the excited state population $\left(\rho_{ee}^{s}\right)_r$ depends on $\overline{\delta}_j$ and $\omega_{rep}$ only through the ratios $\overline{\delta}_j/\omega_{rep}$ and  $\Delta_{12}/\omega_{rep}$, see Eq.(\ref{Eq:DMQSSsimplea}).

Notice that the profile in Fig.\ref{Fig:QSSDM}(b) is asymmetric, while the one on Fig.\ref{Fig:QSSDM} (a) is symmetric with respect to the integer values of $\kappa$. To explain this assymetry we parameterize $\bar{\eta}_1$ and $\kappa$ as  $\bar{\eta}_1=2\pi n_a+\delta\eta$, $\kappa \rightarrow n_b+\delta\kappa$, where free parameters $\eta$ and $\kappa$ are constrained as $0\leqslant\delta\eta<2\pi$ and $0\leqslant\delta\kappa<1$.
When the $n_a$-th harmonic is resonant with the frequency $\omega_{eg_1}$, $\delta\eta=0$.  Then different values of  $\kappa=n_b\pm\delta\kappa$ correspond to the frequency $\omega_{eg_2}$ being red(blue) detuned with respect to the $(n_a-n_b)$-th mode by  $\delta \kappa \omega_{rep}$. Corresponding values of residual detunning $\overline{\delta}_2$ is $\overline{\delta}_2=\pm\delta\kappa\omega_{rep}$ if $\delta\kappa<1/2$ and $\mp(1-\delta\kappa)\omega_{rep}$ if $\delta\kappa>1/2$.  Flipping the sign of $\overline{\delta}_2$ (at $\bar{\delta}_1=0$) does not affect time-evolution of the system, causing the dependence of $\left(\rho_{ee}^{s}\right)_r$ on $\kappa$ (Fig.~\ref{Fig:QSSDM}(a)) at $\bar{\eta}_1=0$ to be symmetrical with respect to $\kappa=n$, $n=0,1..$.

If $\overline{\eta}_{1}$ differs from the integer multiple of $2\pi$, $\overline{\eta}_{1}=2\pi n_a+\delta\eta$, (where $\delta\eta<2\pi$), then the $n_a$-th FC harmonic is detuned from the frequency $\omega_{eg_1}$ by $\frac{\delta\eta}{2\pi}\omega_{rep}$. At $\kappa=n_b\pm\delta\kappa$ both frequencies $\omega_{eg_j}$ generally do not match any of the FC modes.
Different values of $\kappa=n_a\pm\delta\kappa$ correspond to different detunings $\overline{\delta}_2$ at fixed value of $\overline{\delta}_1$, causing the dependence (Fig.~\ref{Fig:QSSDM}(b)) of $\left(\rho^s_{ee}\right)_r$ on $\kappa$ at $\bar{\eta}_1=-\pi/10$ to be asymmetric. However the ``translational'' symmetry with respect to the shift $\kappa=\kappa\pm n$, $n=0,1..$ still remains.

As we showed for fixed $\kappa$ the optimal value of residual detuning, is $\left(\overline{\delta}_1\right)_{opt}=-mod(\pi\kappa,2\pi)/T$.
Now we would like to vary $\kappa$ in order to optimize $\rho_{ee}$ further. One can find that this optimal value of $\kappa=\kappa^{opt}$  can be expressed as
\begin{equation}
\kappa^{opt}=\frac1\pi\arccos(x),\label{Eq:kappaopt}
\end{equation}
where $x$ is a root of the following algebraic equation:

\begin{equation}
16 x^4 \cos ^4\frac{\Theta }{4}-32 x \cosh\frac{\gamma T }{2} \sin ^4\frac{\Theta }{4}+16 \cos\frac{\Theta }{2}-2 x^2 \left(4 \cos \frac{\Theta }{2}+3 \cos\Theta+9\right)=0. \label{Eq:findkappa}
\end{equation}
At fixed values of the decay rate $\gamma$, pulse area $\Theta$, and the frequency gap between the two ground states $\Delta_{12}$, the equation (\ref{Eq:findkappa}) is a self-consistent equation for $T$.

\subsection{Maximum post-pulse excited state population in the quasi-steady-state regime}
In previous subsection we found that the maximum of the post-pulse excited state population $(\rho_{ee}^s)_r $  is reached at
optimal residual detunings $\bar{\delta}_1=-\bar{\delta}_2=-mod(\kappa^{opt}/2, 1)/T$ and optimal parameter $\kappa=\kappa^{opt}$ determined by Eq. (\ref{Eq:kappaopt}).

\begin{figure}[h]
\begin{center}
\includegraphics*[width=3.2in]{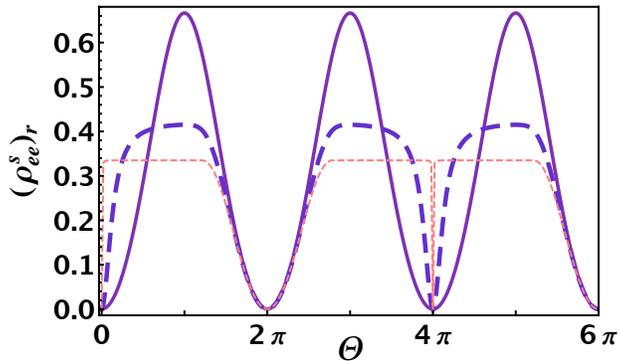}
\end{center}
\caption
{ The dependence of the quasi-steady-state values of the post-pulse excited state population $\left(\rho_{ee}^s\right)_r$ on effective single pulse area $\Theta$ at different values of $\mu=\gamma T$: $\mu=10$ (dashed pink line),
$\mu=1/2$  (dashd blue  line),  $\mu=1/100$  (solid purple  line)
and optimal parameters $\bar{\eta}_1=-\kappa^{opt}/2$,  where $\kappa^{opt}$  is obtained from Eq. (\ref{Eq:kappaopt}).
\label{Fig: maxGSSPop}} \end{figure}

Now we would like to vary the pulse area $\Theta$ to optimize this maximum.
In Fig.~\ref{Fig: maxGSSPop} we plot the dependence of $\left(\rho_{ee}^s\right)_r$ on the effective single pulse area $\Theta$. Different curves correspond to different values of parameter $\mu=\gamma T$. The values of $\left(\rho_{ee}^s\right)_r$  were calculated at the
optimal value of $\kappa$, determined by Eq. (\ref{Eq:findkappa}) for each $\Theta$ and $\mu=\gamma T$.

From Fig.~\ref{Fig: maxGSSPop} we see that the maximum values of the  excited state population are attained at $\Theta=\pi+2\pi n$.
Substituting $\Theta=\pi$ in  the equation (\ref{Eq:findkappa}),  one finds:
\begin{equation}
\kappa^{opt}_{\Theta=\pi}=\frac12+n,\quad n=0,1...
\end{equation}
For these values of $\kappa$ and $\Theta$ the excited state population and the fractional momentum kick are:
\begin{eqnarray}
(\rho_{ee}^s)_r(\Theta=\pi,\bar{\eta}_1=-\pi\kappa,\kappa=\frac12)&=&\frac{1}{3} e^{\gamma T /2} /\cosh(\gamma T/2)\\
\end{eqnarray}
The spectral resolution of the excited state population vanishes as $\Theta\rightarrow\pi$ and $\kappa=\kappa_{opt}$.

The maximum of $\left(\rho_{ee}^s\right)_r$ in three-level $\Lambda$-system (with $b_1=b_2=1/2$, $\theta_1=\theta_2=\sqrt{2}\pi$) is reached at $\gamma T\gg1$.  Its value is $2/3$ that is different from the case of two-level system,  where the maximum excited state population in the quasi-steady state regime is $1$, as it was shown in our previous work \cite{IliAhmDer11}.

In case of unequal pulse areas $\theta_1\neq\theta_2$ and branching ratios $b_1\neq b_2$, at $\Theta=\pi$ and $\gamma T\gg1$, the three-level $\Lambda$-system, which is initially in the ground state $|g_1\rangle$, eventually reaches the QSS  with post-pulse excited state population expressed as
\begin{eqnarray}
\left(\rho_{ee}^s\right)_r\left(\Theta=\pi,\gamma T\gg1\right)=\frac{2\sin^2\pi\kappa\sin^2(2\chi)}{(b_1- b_2) \cos (2\chi )+1+2\sin^2\pi\kappa\sin^2 (2\chi )}.\label{Eq:kickmax}
\end{eqnarray}
If the branching ratios vary as $b_1=\sin^2\chi=\frac{\theta_1^2}{\Theta^2}$, $b_2=\cos^2\chi=\frac{\theta_2^2}{\Theta^2}$,  the Eq.  (\ref{Eq:kickmax}) does not depend on the value of $\chi$,
\begin{eqnarray}
\left(\rho_{ee}^s\right)_r\left(\Theta=\pi,\gamma T\gg1,b_1=\sin^2\chi\right)=\frac{2\sin^2\pi\kappa}{2\sin^2\pi\kappa+1}.\label{Eq:kickmaxreduced}
\end{eqnarray}

The maximum value of $\left(\rho_{ee}^s\right)_r$ in this case is reached at $\kappa=1/2+n$, $n=0,1..$ and is equal to 2/3.  
In case if  $\theta_1=0$, the system which starts in the ground state $|g_1\rangle$ obviously stays unperturbed.  In case if $\theta_1=\pi$,  $\chi=\pi/2$, $b_1=1$ the system is reduced to a pair of coupled levels $|g_1\rangle$ and $|e\rangle$. Then, the maximum population inversion and fractional momentum kick are equal to $1$.


\section{Conclusion \label{Sec: Conclsn}}

In this paper we studied the dynamics of a three-level $\Lambda$-type system driven by a train of ultra-short laser pulses.
General analytic expressions for time-evolution of the density matrix were obtained.
Several regimes of system dynamics can be realized depending on the train parameters.

In particular, when the two-photon resonance condition $\Delta_{12}/\omega_{rep}=0,1..$ is fulfilled, the system evolves into a stationary``dark'' state where it becomes transparent to the pulses. 

In the limiting case when the total pulse area is a multiple of $2\pi$, the ``dark'' state is avoided. In this case, regardless of the pulse repetition rate and the decay rate of the excited state the post-pulse excited state population vanishes.  The system oscillates between the two ground states avoiding populating the excited state alltogether. In a special case of equal pulse areas the entire population can be transfered from one ground state to another by a single $\Theta=2\pi$ pulse.  

At finite excited state decay rates, the system eventually reaches the quasi-steady-state regime which is similar to the saturated regime in a system of two kicked coupled damped pendula. In the QSS regime the radiative-decay-induced drop in the population following a given pulse is fully restored by the subsequent pulse. 

We derived analytical expression for the density matrix in the QSS regime, neglecting the decay during the pulse. The post-pulse excited state population has a periodic dependence on the Doppler shifted phase offset between the subsequent pulses. 
This periodic pattern reflects the frequency comb spectrum and strongly depends on the ratio between the pulse repetition period and the excited state lifetime.

In a particular case when the pulse repetition period is much longer then the excited state lifetime, the interference between subsequent pulses vanishes and the spectral dependence of the excited state population mimics the spectral profile of an individual pulse.

In the opposite case when the excited state lifetime is much longer then the pulse repetition period and the single pulse area is small $\Theta\sim\gamma T$ the pulse train acts on a system as a collection of narrow-band CW lasers with individual frequencies corresponding to  different FC modes. 

At a given pulse area the maximum of excited state population  is reached at some optimal ratio of residual detunings between the frequencies of the two allowed transitions and the nearest FC teeth. This optimal value depends on the effective single pulse area, branching ratios and the ratio of individual pulse areas.  At equal branching ratios and equal pulse areas the optimal residual detunings have the same absolute value and opposite sign.
The single pulse area corresponding to the maximum population inversion is equal to $\pi$.
In case when the ratio of individual pulse areas are determined by the ratio of corresponding dipole matrix elements only, the absolute maximum of the QSS population inversion in the saturation regime (reached at $\Theta=\pi$ and $\gamma T\gg1$)  does not depend on the ratio of these dipole matrix elements and is equal to $2/3$.  In this case the optimal residual detunings are $\bar{\delta}_{1,2}=\pm\frac{\omega_{rep}}4$. This result is different from the case of two-level system, where the maximum population inversion in the saturation regime was equal to $1$. 


\appendix

\section{Density matrix in the saturation regime}
\label{App:rhoSaturation}
Here we derive the value of the density-matrixreached in the saturation (quasi-steady-state) regime.
The pre- and post-pulse elements of the density matrix at the $N^\mathrm{th}$ pulse are related by Eq.(\ref{Eq:Rhoaccross})

\begin{equation}
\left(  \rho^{N}\right)
_{r}=\mathbf{A}_N~\left( \rho^{N}\right)
_{l}~\mathbf{A}_N^\dagger,
\label{Eq:RhoaccrossApp}%
\end{equation}

Introducing the unitary transformation

\begin{equation}
(\check{\rho})^N_{r}=u_{N}^\dag(\rho)^N_{l,r} u_{N},\label{bebe}
\end{equation}
where

\begin{equation}
u_N=\left(
\begin{matrix}e^{i\frac{\eta_1(t_{N})+\eta_2(t_{N})}2}&0&0\\0&e^{i\frac{\eta_2(t_{N})-\eta_1(t_{N})}2}&0\\0&0&e^{i\frac{\eta_1(t_{N})-\eta_2(t_{N})}2}
\end{matrix}
\right),
\end{equation}
one can rewrite (\ref{Eq:RhoaccrossApp}) as:

\begin{equation}
\left(\check{\rho}^{N}\right)
_{r}=\mathbf{A}_0~\left( \check{\rho}^{N}\right)
_{l}~\mathbf{A}_0^\dagger.
\label{Eq:RhoaccrossAppU}%
\end{equation}
The quasi-steady-state density matrix $\left( \check{\rho}^{N}\right)_r=\left( \check{\rho}^s\right)_r$ can be obtained from the system of linear equations
\begin{equation}
\left( \check{\rho}^{N}\right)_r=\left( \check{\rho}^{N-1}\right)_r.
\end{equation}
The general equation for the post-pulse  excited state population can be expressed then as

\begin{equation}
\left(\rho_{ee}^s\right)_r\frac{4 e^{\gamma T } \sin ^4\left(\frac{\theta }{2}\right) \sin ^2\left(\pi\kappa\right) \sinh ^2\left(\frac{\gamma T}{2}\right) \sin ^2(2 \chi )}{D},\label{Eq:rgen}
\end{equation}
where

\begin{equation}
D= Det \left(\lbrace D_1,D_2, D_3,D_4, D_5,D_6, D_7,D_8\rbrace\right)
\end{equation}

\begin{eqnarray}
D_1&=&\lbrace 1-e^{\gamma T}-(1+\cos^2\chi)\sin^2\frac{\Theta}2,-\sin^2\frac{\Theta}2\cos 2\chi ,0,0,\sin^2\frac{\Theta}2\sin2\chi,\nonumber\\
&&\sin\Theta\sin\chi,\sin\Theta\cos\chi,0\rbrace\nonumber\\
D_2&=&\lbrace2b_1e^{\frac{\gamma T}2}\sinh\frac{\gamma T}2+4\sin^2\chi\sin^2\frac{\Theta}4(1-\sin^2\frac{\Theta}4(1+\cos^2\chi)),\nonumber\\
&& -4 \sin ^2\frac{\Theta }{4}\sin ^2\chi  \left(\sin ^2\frac{\Theta }{4}\cos 2 \chi +1\right),0,0,-2 \sin ^2\frac{\Theta }{4}\sin 2\chi \left(1-2 \sin ^2\frac{\Theta }{4}\sin ^2\chi\right),\nonumber\\
&&-2 \sin\frac{\Theta }{2}\sin\chi\left(1-2 \sin ^2\frac{\Theta }{4}\sin ^2\chi\right),2 \sin ^2\frac{\Theta }{4}\sin\frac{\Theta }{2}\sin\chi \sin2\chi, 0 \rbrace,\nonumber\\
D_3&=&\lbrace 0,0, 1-e^{\frac{\gamma T} 2} \cos\bar{\eta}_1-2 \sin ^2\frac{\Theta }{4}\cos ^2\chi,
\sin ^2\frac{\Theta }{4} \sin2 \chi ,0,
e^{\frac{\gamma T}2} \sin \bar{\eta}_1,0,
\sin\frac{\Theta }{2}\cos\chi\rbrace,\nonumber\\
D_4&=&\lbrace 0,0, \sin ^2\frac{\Theta }{4}\sin 2 \chi,
1-e^{\frac{\gamma T}2} \cos (\bar{\eta}_1 +2 \pi  \kappa )-2 \sin^2\frac{\Theta }{4}\sin ^2\chi,0,0,
e^{\frac{\gamma T}2} \sin (\bar{\eta}_1 +2\pi\kappa ),\nonumber\\ &&-\sin\frac{\Theta }{2}\sin\chi\rbrace,\nonumber\\
D_5&=&\lbrace \sin^2\frac{\Theta}4\sin2\chi(3\cos^2\frac{\Theta}4-\sin^2\frac{\Theta}4\cos2\chi),-\sin ^4\frac{\Theta }{4}\sin 4\chi, 0, 0,\nonumber\\&&
-\sin ^4\frac{\Theta }{4} \cos 4 \chi +\cos ^4\frac{\Theta }{4}-\cos2\pi\kappa,
-\cos\chi \sin\frac{\Theta }{2} \left(1-4 \sin ^2\chi  \sin^2 \frac{\Theta}4\right),\nonumber\\
&&\sin\frac{\Theta }{2}\sin\chi\left( 4\cos^2\chi\sin^2\frac{\Theta}4-1\right),
\sin (2\pi\kappa )\rbrace,\nonumber\\
D_6&=&\lbrace \sin\chi \sin\frac{\Theta}2( \cos^2\chi- (1+\cos^2\chi)\cos\frac{\Theta}2),\sin \frac{\Theta }{2}\sin\chi\left(2 \sin ^2\frac{\Theta }{4}\cos 2\chi+1\right),
-e^{\frac{\gamma T} 2} \sin \bar{\eta}_1,0,\nonumber\\&&
\sin\frac{\Theta }{2}\cos \chi \left(1-4 \sin ^2\frac{\Theta }{4}\sin ^2\chi\right),-e^{\frac{\gamma T}2} \cos\bar{\eta}_1+\cos\frac{\Theta }{2} \cos ^2\chi +\cos\Theta  \sin ^2\chi ,\nonumber\\
&&\left(\cos\Theta-\cos\frac{\Theta }{2}\right) \frac{\sin 2 \chi}2, 0\rbrace,\nonumber
\end{eqnarray}
\begin{eqnarray}
D_7&=&\lbrace 2\cos\chi \sin\frac{\Theta}2(\sin^2\frac{\Theta}4(1+\cos^2\chi)-1),-\cos\chi \sin \frac{\Theta }{2}\left(1-2 \sin ^2\frac{\Theta }{4}\cos 2\chi\right) , 0,\nonumber\\&&
-e^{\frac{\gamma T}2} \sin (\bar{\eta}_1 +2 \pi  \kappa ),\sin \frac{\Theta }{2}\sin\chi\left(1-4 \sin ^2\frac{\Theta }{4}\cos ^2\chi\right),
-\sin ^2\frac{\Theta }{4}\sin 2 \chi\left(2 \cos\frac{\Theta }{2}+1\right) ,\nonumber\\&&-e^{\frac{\gamma T}2} \cos (\bar{\eta}_1 +2\pi \kappa )+\left(\cos \frac{\Theta }{2}-\cos\Theta\right) \sin ^2\chi+\cos\Theta,0\rbrace,\nonumber\\
D_8&=&\lbrace 0, 0, -\cos\chi\sin\frac{\Theta }{2},
\sin \frac{\Theta }{2}\sin \chi,
-\sin (2 \pi  \kappa) ,0,0,
\cos\frac{\Theta }{2}-\cos (2\pi\kappa )\rbrace\nonumber
\end{eqnarray}

In some limiting cases the general equation (\ref{Eq:rgen}) can be simplified further.

\begin{enumerate}

\item[(a)]{At $\theta_{1}=\theta_{2}$:

\begin{equation}
\left(\rho_{ee}^s\right)_r=\frac{2 e^{\frac{\gamma T}{2}}}{D^a} \sin ^2\left(\pi\kappa\right) \sin^2\frac{\Theta}2,\label{Eq:DMQSSbranch}
\end{equation}
where
\begin{eqnarray}\label{Eq:D}
D^a&=&\left(b_1 \cos\overline{\eta}_1+b_2 \cos\overline{\eta}_2\right) \left(4 \cos
\frac{\Theta}2-\sin^2\frac{\Theta}2-2\cos(2\pi\kappa)\cos^4\frac{\Theta
}{4}\right)-\nonumber\\ & & -2 \left(b_2 \cos\overline{\eta}_1+b_1\cos\overline{\eta}_2\right)
\left(\sin ^4\frac{\Theta }{4}+\cos
^2\frac{\Theta}2\right)+\nonumber \\
&&+2\sin(2\pi\kappa)(b_2\sin\overline{\eta}_2+b_1 \sin\overline{\eta}_1)\cos^4\frac{\Theta
}{4}-\nonumber \\&&-2 \cosh\frac{\gamma T}{2} \left(\sin^4\frac{\Theta }{4}+\cos ^2\frac{\Theta }{4}
\sin ^2\left(\pi\kappa\right)\right).
\end{eqnarray}

}

\item[(b)]{At $b_1=\sin^2\chi$, $b_2=\cos^2\chi$:

\begin{eqnarray}
\left(\rho_{ee}^s\right)_r&=&8e^{\frac{\gamma T}2 } \sin^2\frac{\Theta}2\sin ^2\pi\kappa/D^b\nonumber\\
D^b&=&8 \cos 2\chi\left(4 \sin^4\frac{\Theta }{4}+\sin ^2\frac{\Theta }{2} \cos 2\pi\kappa\right)\sin\pi\kappa\sin \left(\bar{\eta}_1 +\pi\kappa\right)+\nonumber\\&&
+\cos\pi\kappa\cos \left(\bar{\eta}_1 +\pi\kappa\right)\left(4 \cos\frac{\Theta }{2}\left(\cos2\pi\kappa-5\right)+(\cos\Theta+3) (3 \cos2\pi\kappa+1)-\right.\nonumber\\&&
\left. -16 \sin ^4\frac{\Theta }{4}\sin ^2\pi\kappa\cos (4 \chi )\right)-\nonumber\\&&
-4\cosh\left(\frac{\gamma T}2\right)\left(4 \cos^2\frac{\Theta }{4}\cos2\pi\kappa+2 \cos\frac{\Theta }{2}-\cos\Theta-5\right).
\end{eqnarray}
}

\item[(c)]{At  $\Theta=\pi$:

\begin{eqnarray}
\left(\rho_{ee}^s\right)_r&=&16e^{\frac{\gamma T}2 } \sin ^2\pi\kappa\sinh (\frac{\gamma T }2) \sin ^22 \chi /D_c\nonumber\\
D_{c}&=&\sinh\left( \frac{\gamma T}2 \right)\left(\cos\bar{\eta}_1\left(2 (2 (b_1-b_2+2) \cos 2\pi \kappa + b_1+5 b_2 \cos 4\pi \kappa  +1)+ \right.\right.\nonumber\\
&&+\cos 2\chi (2 (8b_1 - 1)\cos 2\pi\kappa +
b_1- 15b_2\cos 4\pi \kappa  + 1)+8 \sin ^2\left(\pi \kappa \right) \cos 4\chi (b_1-2-\nonumber\\&&-3 b_2 \cos2\pi \kappa )+
  \left.  4 \sin ^2\pi \kappa \cos 6 \chi
   (b_2 \cos2\pi \kappa -b_1)\right)-8 \sin \bar{\eta}_1 \left(4 b_2 \sin 4\pi\kappa  \sin^6 \chi+\right.\nonumber\\
&& \left. \left. +\sin 2\pi \kappa  \cos^2 \chi (b_2 (\cos 4\chi+3)-4 \cos2 \chi)\right)\right)+\nonumber\\
&&+4 \sinh (\gamma T) \left(2 (b_1-b_2) \cos 2\chi -2 \cos 2\pi\kappa \sin ^2 2\chi -\cos 4\chi+3\right)
   \end{eqnarray}

}

\item[(d)]{At $\Theta=\pi$, and $b_1=\sin^2\chi$, $b_2=\cos^2\chi$:

\begin{eqnarray}
\left(\rho_{ee}^s\right)_r&=&16e^{\frac{\gamma T}2 } \sin ^2\pi\kappa /D^d\nonumber\\
D^d&=&\cos \bar{\eta}_1 (8 \cos2\pi\kappa+4 \cos 2\chi-\cos (4 \chi )+5)-16 (\cos2\pi\kappa-2) \cosh \left(\frac{\gamma T }{2}\right)+\nonumber\\
&&2 \sin
   \bar{\eta}_1 \sin 2\pi\kappa(4 \cos 2\chi+\cos (4 \chi )-1)+ 8 \sin ^4\chi \cos
   (\bar{\eta}_1 +4\pi\kappa).
\end{eqnarray}
}

\end{enumerate}

\section*{Acknowledgments}
We would like to thank Mahmoud Ahmad for discussions.
This work was supported in part by the NSF and ARO.


\end{document}